# Properties and nanotechnological applications of nanotubes

Yu E Lozovik, A M Popov

**Abstract**
Possible applications of carbon nanotubes in nanoelectromechanical systems (NEMSs) based on relative motion and interaction of nanotubes walls are considered for wide set of NEMS. This set includes nanomotors, nanoactuator, nanorelay, variable nanoresistor, gigahertz oscillator, nanothermometer and so on. The nanotube properties and theory for relative motion and interaction of nanotubes walls is described. The principal schemes, operation principles, and methods of actuation of the considered NEMS are considered. The development of nanotechnological techniques which are necessary for production of nanotube-baseb NEMS are discussed.

## 1. Introduction

The last decade has seen considerable progress in nanomechanics. In particular, the feasibility of manipulating nanometer-sized objects has been demonstrated [1] and basic designs for nanoelectromechanical systems (NEMSs) which can realize controllable motion of nanoobjects have been considered [2]. The search for nanoobjects that may be employed as the movable elements of NEMSs is currently a topical problem. The possibility of arbitrary [3, 4] relative motion of walls in multi-walled carbon nanotubes [7] that is controllable with an atomic-force microscope [5, 6] and the extraordinary elastic properties of carbon nanotube walls [8–12] make them challenging candidates for NEMS elements. Several nanomechanisms have been suggested, which are based on the relative sliding or rotation of nanotube walls: a nanobearing [13], a nanogear [14], a nanoswitch [15], a nanorelay [16, 17], a gigahertz oscillator [18, 19], and a Brownian nanomotor [20, 21]. The unique electronic properties of carbon nanotubes [22] find use in experimentally realized electronic nanodevices like nanotransistors [23], nanodiodes [24], current nanomodulators [25], etc. Furthermore, nanomotors reliant on the relative rotation of carbon nanotube walls have been made recently [26, 27]. In the nanoresistor, nanorelay, and nanomotor mentioned above, the nanotube walls are simultaneously movable NEMS elements and elements of an electric circuit. This family of carbon nanotube applications in NEMSs embraced a new unique and promising application: specifically, it was shown that a double-walled carbon nanotube (DWNT) could be a pair with an effective `screw thread' [28–32]. In this connection, basic designs were proposed for NEMSs based on carbon nanotubes whose structure comprises a `nanobolt-nanonut' pair: a `nanodrill' for nanolocal surface modification [28–30, 32] and a nanoactuator for the transformation of a force directed along the nanotube axis to the relative rotation of its walls [17, 32, 33]. According to calculations, the nanotube conductivity significantly changes under a minor relative displacement (by a fraction of an angström) of its walls along the nanotube axis [34–36] or a relative rotation of the walls [37]. We came up with the idea of employing this nanotube property in such NEMSs as a variable nanoresistor [28–30], tension nanosensor [38], and electromechanical nanothermometer [17, 39].

Carbon nanotubes are promising candidates for utilizing in NEMSs based not only on the relative motion of nanotube walls, but also on nanotube bending. Of these NEMSs mention should primarily be made of nanooscillators in which measurements of the frequency of transverse nanotube modes are employed for measuring the properties of a system. For instance, by measuring the vibration frequency of a nanotube fixed at one end it was possible to determine the mass of a nanoparticle and even of a single molecule located at the other end of the nanotube [11]; in addition, a nanodevice was realized which allowed, by measuring the vibration frequency of a nanotube fixed at both ends, measuring the force acting on the nanotube with an accuracy of $5 \times 10^{-18}$ N [40]. Furthermore, nanotweezers were made, which capture nanoparticles and manipulate them by bending the ends of two nanotubes via electrostatic forces [41, 42], as was a memory cell of two intersecting nanotubes [43].

This report is concerned with NEMSs based on carbon nanotubes with movable walls. The properties and structure of nanotubes are described in Section 2. The interaction and relative motion of nanotube walls are considered in Section 3. Ways of controlling the motion of nanotube walls and NEMS operating modes are discussed in Section 4. Section 5 is dedicated to the basic designs and operating principles of carbon nanotube-based NEMSs. Finally, considered in Section 6 is the development of nanotechnology methods which may be employed to fabricate carbon nanotube-based NEMSs.

## 2. Structure and properties of nanotubes

### 2.1 Structure of single-walled nanotubes

A single-walled carbon nanotube (SWNT) may be represented as a single graphite plane rolled up (graphene). The SWNT structure is defined by the pair of integers ($n,m$) – chirality indices, which are the coordinates of the grapheme lattice vector $\mathbf{c}=n\mathbf{a}_1+m\mathbf{a}_2$ (where $\mathbf{a}_1$ and $\mathbf{a}_2$ are the unit vectors of the graphene plane). The segment corresponding to the vector c becomes, on rolling up a graphene plane fragment into an SWNT, its circumference (Fig. 1) [22, 44]. The wall radius $R$ is defined by the expression

$$R = \frac{|c|}{2\pi} = \frac{a_0\sqrt{n^2+mn+m^2}}{2\pi} \quad (1)$$

where $a_0$=0.246 nm is the length of the unit vector of the graphene plane. The SWNT radius typically is on the order of several nanometers; the shortest SWNT radius obtained is about 0.4 nm. The SWNT unit cell length is expressed as

$$b = \frac{\sqrt{3}a_0\sqrt{n^2+mn+m^2}}{\mathrm{GCD}(2m+n, n+2m)} \quad (2)$$

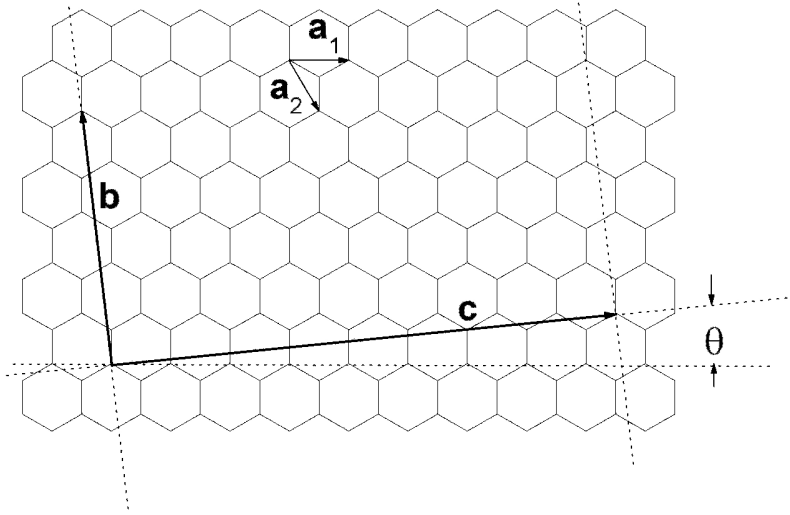

**Figure 1**. Graphite plane (graphene) rolled into a nanotube wall; $\mathbf{a}_1$ and $\mathbf{a}_2$ are the unit vectors of the graphite plane. The wall is unambiguously defined by the vector $\mathbf{c}$: the length of wall circumference is equal to $|\mathbf{c}|$. The modulus of the vector $\mathbf{b}$ defines the unit cell length, and θ is the wall chirality angle.

where GCD ($u,w$) is the greatest common divisor of the integers $u$ and $w$. The chirality angle $\theta$ is defined as the angle between the vectors $\mathbf{a}_1$ and $\mathbf{c}$:

$$\theta = -\arccos\frac{2n+m}{\sqrt{n^2+m^2+mn}} \quad (3)$$

Since the graphite plane is sixth-order symmetrical, only the SWNTs defined by vectors $\mathbf{c}$ that lie within an angle of 60°, i.e., with $m > 0$ and $n > 0$, are nonequivalent. Single-walled carbon nanotubes with the chirality indices ($n, n$) — an `armchair' — and ($n, 0$) — a `zigzag' — are nonchiral, whereas all the rest of the SWNTs are chiral. In this case, SWNTs with chirality indices ($n, m$) and ($m, n$) are mirror-symmetric, i.e., are structurally left-helical for $n > m$, and right-helical for $m > n$.

### 2.2 Electronic properties of nanotubes

As is well known, a graphene sheet possesses an electronic structure with a linear dispersion law in the conduction and valence bands (i.e., the effective masses of electron and hole excitations are equal to zero) and a zero energy gap. Rolling up the graphene sheet results in band confinement (addition), with the effect that there appear `transverse quantization' subbands, depending on chirality indices. As a result, the SWNTs with chirality indices ($m$, $n$) are metallic, when the condition $m+n=3q$ is satisfied, where $q$ is an integer [22], and are semiconducting in the remaining cases.

### 2.3 Elastic properties of nanotubes

Both SWNTs [45, 46] and multi-walled carbon nanotubes (MWNTs) up to 30 μm in length [47] and containing several dozen walls [7] have been obtained. Multi-walled carbon nanotubes have the structure of nested coaxial walls, similar to the graphite structure, with an interwall distance of about 0.34 nm [7], which is close to the spacing between graphite layers. That is why graphite and nanotubes possess similar elastic properties. In particular, the experimental elastic modulus for graphite compression along the layers amounts to 1.24 TPa [48], which is approximately five times greater than the elastic modulus for steel.

The elastic properties of nanotubes, including their plastic deformation under a heavy load, were considered at length in the review [49], and therefore in this report we restrict ourselves to a brief enumeration of the main results. Numerous experiments have yielded a value of about 1 TPa for the Young modulus. Specifically, measurements of the force required for bending an MWNT, an SWNT, and SWNT bundles yielded Young moduli of 1.28±0.59 TPa [9], 1.2 TPa [50], and 1.18±0.4 TPa [51], respectively. Measurements of the force required for the extension of nanotubes yield values of the Young modulus of 0.27–0.95 TPa for MWNTs [3] and 0.32–1.47 TPa for SWNT bundles [52]. Young moduli of 1.8±1.4 TPa for MWNTs [8] and 1.3+0.6/–0.4 TPa for SWNT [10] were derived from an analysis of the thermal vibrations of nanotubes fixed at one end. Investigation into the electromechanical vibrations of MWNTs fixed at one end shows that their flexural modulus increases from 0.1 to 1 TPa as the nanotube diameter decreases [11].

The experimental values of the Young modulus agree well with the theoretical estimates. Values of the Young modulus for different SWNTs, obtained by way of calculations based on the density functional theory, range between 0.94 and 1.09 TPa [53–57].Values of the Young modulus close to 1 TPa were also obtained via semiempirical methods [58–61].

### 3. Interaction and relative motion of nanotube walls

The force causing the relative motion of nanotube walls is defined by the expression
$$F = F_\mathrm{f} + F_\mathrm{W} = \pi D \tau L_\mathrm{ov} + \gamma \pi D + F_\mathrm{e} \qquad (4)$$
where $F_\mathrm{f}$ is the force of static and dynamic friction, $F_\mathrm{W}$ is the van der Waals force which retracts the inner wall into the outer one in the telescopic extension of the former, $D$ is the diameter of the movable wall, $L_\mathrm{ov}$ is the wall overlap length, $\tau$ is the wall shear modulus which includes the dynamic and static shear moduli, $\gamma$ is the interwall interaction energy per unit area of the wall overlap surface, and $F_\mathrm{e}$ is the static friction force related to the interaction between the edge of one wall and the surface of the other one. By way of investigations into the interwall interaction force with the aid of atomic-force microscopy [6] it was determined that the interwall interaction energy g ranges from 22 to 33 meV per atom for different nanotubes. This value is consistent with the estimate obtained from the geometrical characteristics of flattened nanotubes: $\gamma=35\pm10$ meV per atom [62], and with the data calculated by the method of density functional for different nanotubes: $\gamma=18$–21 meV per atom [63], and $\gamma=23$–24 meV per atom [12]. This interwall interaction energy corresponds to values of the van der Waals force from 0.5 to 20 nN, depending on the diameter of the movable wall. Forces in this range are typical for atomic-force microscopy (see the review [64]), and the relative motion of walls may therefore be controlled using a nanomanipulator.

The above experiments [6] revealed no energy dissipation in the relative motion of walls and showed that the force required for the telescopic extension of the inner wall is independent of the wall overlap length. These results yield upper estimates for the dynamic and static shear moduli of 0.05 and 0.04 MPa, respectively. The experimental value of the shear modulus for graphite layers is about 1 MPa [65]. The substantially lower value of the shear modulus for nanotube walls in comparison with the shear modulus for graphite layers is explained as follows: graphite layers are always commensurate, while nanotube walls may be both commensurate and incommensurate.

The walls of a nanotube are commensurate when the ratio $b_1/b_2$ between the unit cell lengths is a rational fraction. The characteristics of the interaction and the relative motion of commensurate and incommensurate nanotube walls are theoretically analyzed below by the example of DWNTs. In the case of commensurate walls, a DWNT is a one-dimensional crystal with the unit cell length equal to the least common multiple of the unit cell lengths of the walls.

Recently, a classification scheme was proposed for DWNTs with commensurate walls [66, 67]. According to this scheme, these double-walled nanotubes make up families. All nanotubes of each family have equal unit cell lengths, interwall distances, and chirality angles for the inner and outer walls. This classification is simultaneously a classification for the possible pairs of commensurate neighboring MWNT walls and may be employed for the selection of such pairs which show promise for NEMS (for instance, for the selection of neighboring walls in a nanoactuator [33]). A table comprising the complete list of all possible DWNT families with commensurate walls is given in Ref. [66].

To investigate the characteristics of relative motion of nanotube walls, it is required to calculate how the interaction energy $U$ for two neighboring walls depends on the coordinates describing the relative position of the walls: the angle $\phi$ of the relative wall rotation about the nanotube axis, and the length $z$ of the relative wall displacement along this axis. The potential relief $U(z,\phi)$ of interwall interaction energy is conveniently visualized by developing the cylindrical surface. The symmetry transformations of the function $U(z,\phi)$ comprise all symmetry transformations of both the walls. There are four combinations of neighboring wall pairs with basically different potential reliefs of their interaction energy: (1) commensurate nonchiral walls; (2) commensurate walls, at least one of which is chiral; (3) incommensurate walls, and (4) commensurate walls, at least one of which is chiral and the other possesses periodically located defects in atomic wall structure.

For a DWNT with commensurate nonchiral walls $((n,n)@(m,m)$ and $(n,0)@(m,0))$, an expression was derived for the expansion of the interwall interaction energy into a Fourier series [68]: $U(z,\phi)$

$$U(z,\phi) = \sum_{M,K(\text{odd})=1}^{\infty} \alpha_K^M \cos\left(\frac{2\pi}{b}Kz\right)\cos\left(\frac{nm}{N}M\phi\right)\sin^2\left(\frac{\pi nm}{2N^2}\right) + \\ + \sum_{M,K(\text{even})=0}^{\infty} \beta_K^M \cos\left(\frac{2\pi}{b}Kz\right)\cos\left(\frac{nm}{N}M\phi\right), \quad (5)$$

where $N$ is the greatest common divisor of $n$ and $m$, and $b$ is the length of an individual DWNT cell. Even terms are always present in expansion (5), while odd terms are present only when both the ratios, $n/N$ and $m/N$, are odd. According to the topological theorem [69], the critical points of the interwall interaction energy $U(z,\phi)$ correspond to the relative positions of the walls for which the second-order symmetry axes $U_2$ of the two walls coincide. The axes $U_2$, perpendicular to the principal wall axis, pass through this axis and the middle of the bond or the center of the hexagon in the wall structure. The elementary cell of the function $U(z,\phi)$ for a DWNT with commensurate nonchiral walls contains four critical points: a minimum, a maximum, and two saddles. The relative wall positions corresponding to the critical points are described in Ref. [12].

The amplitudes of the harmonics in expansion (5) decrease exponentially with harmonic numbers $M$ and $K$ [69–71]. That is why the interwall interaction energy $U(z,\phi)$ for a DWNT with commensurate nonchiral walls may be approximated with only the first two terms of expansion (5):

$$U(z,\varphi) = U_0 - \frac{\Delta U_\varphi}{2}\cos\left(\frac{2mn}{N}\varphi\right) - \frac{\Delta U_z}{2}\cos\left(\frac{4\pi}{b}z\right), \quad (6)$$

where $U_0$ is the average interwall interaction energy, and $\Delta U_\varphi$ and $\Delta U_z$ are the potential barriers for the relative rotation of the walls and for their sliding along the DWNT axis, respectively. In this case, $U_0$, $\Delta U_\varphi$, and $\Delta U_z$ are defined by interpolation from the values of $U(z,\phi)$ at four critical points. Semiempirical calculations show, by the example of several dozen DWNTs with commensurate nonchiral walls, that the energy $U(z,\phi)$ can be approximated by expression (6) correct to about 1% [67]. This result is confirmed by the data of calculations done by the method of density functional for a DWNT $(5,5)@(10,10)$ with an accuracy of about 5% [63].

The majority of DWNTs with commensurate nonchiral walls have incompatible wall rotation symmetries. For this reason, the angular period $\delta_\varphi = 2nm/N$ of the energy $U(z,\phi)$ is short, and the po-tential barrier $\Delta U_\varphi$ for relative wall rotation is low (lower than 0.005 meV per atom according to calculations by the method of density functional [63], and lower than $10^{-11}$ meV per atom according to calculations employing semiempirical potentials [67]). In this case, the interwall interaction energy is approximately defined by the equation

$$U(z,\varphi) \approx U_0 - \frac{\Delta U_z}{2}\cos\left(\frac{4\pi}{b}z\right) \qquad (7)$$

The sole exceptions are provided by DWNT (5,5)@(10,10) [12, 31, 63, 71–73] and DWNT (9,0)@(18,0) [12, 67, 71], which exhibit significant barriers for relative wall rotation. The potential reliefs of the interwall interaction energy $U(z,\phi)$ for DWNTs (5,5)@(10,10) and (6,6)@(11,11), which possess compatible and incompatible wall rotation symmetries, are depicted in Figs 2a and 2b, respectively. We note that expression (5) is valid for all physical quantities which depend on the relative position of non-chiral commensurate DWNT walls, in particular, for the conductivity.

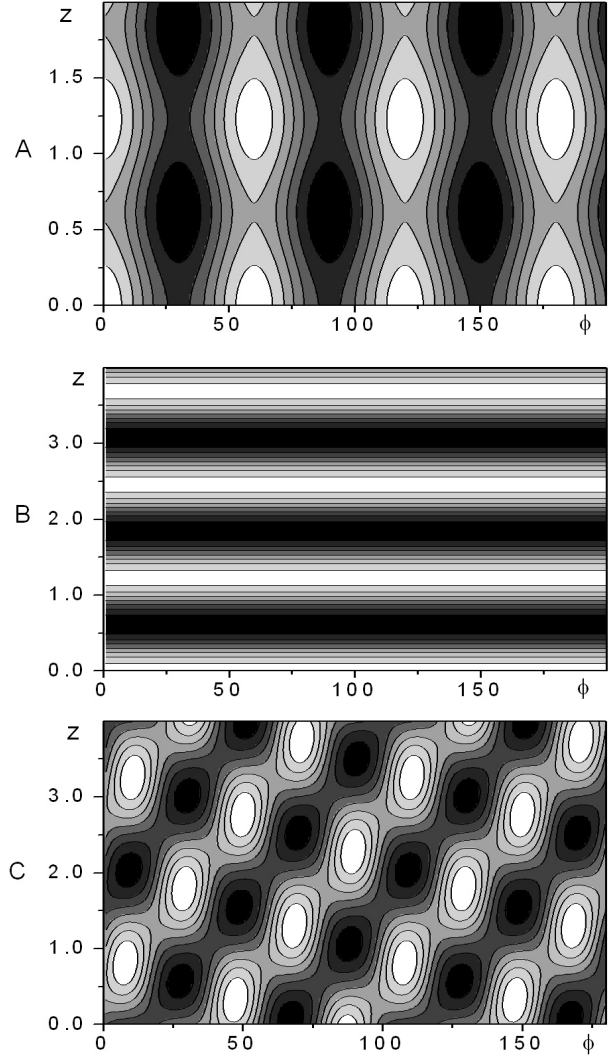

**Figure 2**. Potential reliefs of the interaction energy $U(z,\phi)$. for the walls of a double-walled nanotube as functions of the relative wall displacement $z$ along the nanotube axis and of the angle $\phi$ of relative wall rotation about the axis for (a) DWNT (5,5)@(10,10), (b) DWNT (6,6)@(11,11), and (c) a DWNT with a vacancy in every unit cell of the inner wall. The energy is measured from its minimum. The equipotential lines in Figs 2a, 2b, and 2c are drawn at $4.0\times10^{-3}$, $1.0\times10^{-3}$, and $4.0\times10^{-3}$ meV per atom intervals, respectively.

The potential relief is quite flat for any DWNT with commensurate walls, at least one of which is chiral. This is so because only very high harmonics of the Fourier transform of the interaction energy $U_a$ between an atom of one wall and the entire other wall make contributions to the potential barriers due to the incompatibility of the helical symmetries of the walls [70]. For instance, the barrier for the relative wall rotation of a DWNT (8,2)@(16,4), which was calculated on the basis of the Lenard–Jones potential, is about $5\times10^{-12}$ meV per atom, and this is the only example revealed to date in which the potential barrier for the relative motion of chiral commensurate walls exceeds (!) the limit of calculation accuracy [67]. The heights of other barriers for several dozen DWNTs with chiral commensurate walls considered are lower than the calculation accuracy [67]. A calculation based on the Crespi–Kolmogorov potential [74] also shows that the potential relief for all the defect-free DWNTs under consideration that have commensurate walls, at least one of which is chiral, is extremely flat [71]. This result was borne out by calculations reliant on the method of density functional for a DWNT (8,2)@(16,4) [63].

In the general case, a nanotube wall possesses helical symmetry (see, for instance, Ref. [75]), and therefore the interwall interaction energy $U(z,\phi)$ also possesses helical symmetry. Consequently, the potential relief may have valleys directed along a helix, similar to the thread on the side surface of a bolt. In what follows such potential reliefs are termed reliefs of the thread type. Double-walled carbon nanotubes with thread type potential reliefs may be employed in NEMSs as a nanobolt–nanonut pair. Among the quantitative characteristics of this thread are the potential barriers $E_1$ and $E_2$ for relative DWNT wall motion along the thread line and for twisting-off the thread (motion across the thread), respectively, as well as the threshold forces for setting the walls in relative motion along the thread line and for twisting-off the thread. The thread may be qualitatively characterized not only by potential barriers, but also by their ratio $\beta=E_2/E_1$ which is termed the relative depth of thread [28, 29]. It was nevertheless discovered that in those cases where DWNTs with defect-free walls may have a thread type relief the characteristics of this thread do not permit employing these DWNTs as a nanobolt–nanonut pair in NEMSs. Initially, the thread type relief was found in several DWNTs with incommensurate walls, in this case for simplicity one of the walls being assumed to be ultimately short — equal in length to that of the wall unit cell (normally equal to several nanometers or dozen nanometers) [31]. However, a detailed study revealed that the depth of thread of the potential relief for a DWNT with incommensurate walls is small and therefore the force required for twisting-off the thread is weak for any wall lengths [28, 29]. Moreover, the depth of thread varies significantly with the length of the short movable wall, and the thread type relief may even vanish for some values of the length of this wall [28, 29]. Therefore, a DWNT with incommensurate walls may be employed as a nanobolt–nanonut pair with the desired depth of thread only when the short wall length corresponds to within several angströms to the prescribed one. The feasibility of fabricating such a DWNT is highly conjectural. Since nonchiral walls possess planes of mirror symmetry passing through the principal wall axis [68], the potential relief of these DWNTs also possesses suchlike mirror symmetry planes and, hence, may not be a thread type relief [32, 76]. As mentioned above, for DWNTs with commensurate walls, at least one of which is chiral, the potential relief is rather flattened due to the incompatibility of wall symmetries, and therefore all the potential barriers for the relative motion of walls are extremely low. Furthermore, a detailed consideration of the potential relief symmetries of these DWNTs reveals that most of them may not possess a thread type potential relief at all [32, 76].

It turns out that the presence of defects in an atomic wall structure radically changes the situation. Specifically, due to the occurrence of defects, even the first harmonic of the Fourier transform of the interaction $U_a$ between an atom of one wall and the entire other wall makes a nonzero contribution to the potential barriers. According to several calculations, this results in an increase in the corresponding barriers by 8–10 (!) orders of magnitude [32, 67, 76, 77]. This property of a DWNT with atomic structure defects may be employed to obtain a DWNT with a thread type relief whose characteristics allow using such a DWNT as a nanobolt–nanonut pair in NEMSs. It has been suggested that similar artificial defects should be produced at identical sites in a large number of unit cells of DWNTs with commensurate walls [32, 67, 76, 77]. In this case, any barrier $\Delta U$ for relative motion of walls is defined by the expression $\Delta U$. $\Delta U = \Delta U_u N_u$, where $\Delta U_u$ is the barrier for one nanotube cell, and $N_u$ is the number of cells with defects. Therefore, the `atomic design' of the DWNT structure permits obtaining the nanobolt–nanonut pair with desired (high enough) barrier values which prevent the thread from being twisted-off in the case of sufficiently long nanotubes. A theoretical investigation was made to elucidate how the thread characteristics are affected by a `substituted atom' defect structure and by the potential parameters for the interaction between the atom corresponding to the defect and the atoms of the defect-free wall [32, 76]. It was found that the relative depth of thread $\beta=E_2/E_1$ (which characterizes its quality) depends only slightly on the above quantities and consequently is determined by the structure of the defect-free wall. An example of a thread type potential barrier for a DWNT with commensurate nonchiral walls and periodically sited defects is given in Fig.2b.

# 4. Nanoelectromechanical systems: operating modes and ways of control

## 4.1 Modes of operation of nanoelectromechanical systems

To investigate the feasibility of utilizing the nanobolt–nanonut pair in NEMSs, an analysis was made of the relative motion of coaxial DWNT walls along the helical line of thread [28–30]. (The results of this analysis also apply to the special cases of helical motion — the relative rotation of the walls and their relative sliding along the DWNT axis.) The researchers underwent the analysis of the case where one wall was fixed and the other movable, while the forces acting on the movable wall did not lead to its deformation and displacement of its axis with respect to the axis of the fixed wall. For instance, this is possible when each atom of the movable wall experiences a force $\mathbf{F}_a$ consisting of two components which are the same for every atom: the force $F_z$ aligned with the DWNT axis, and the force $F_l$ aligned with the tangent to the wall circumference, passing through the atom. In the case under consideration, it is easily shown that the motion of the movable wall of mass $M$ relative to the fixed wall is equivalent to the two-dimensional motion of a particle of mass M in the potential field of the interwall interaction energy $U(R\phi,z)$, where $R$ is the movable wall radius, under the action of force $\mathbf{F}=(N_aF_z, N_aF_l)$. In this case, the wall motion along the helical line of the thread is equivalent to the particle motion along a straight line.

As the analysis suggests, when (i) the potential barrier for twisting-off the thread is significantly higher than the barrier for relative motion of walls along the line of thread, $E_2>>E_1$, (ii) the energy of thermal wall motion is significantly lower than the barrier heights for their relative motion, $kT<<E_1,E_2$, and (iii) the force acting on the movable wall is weak, $F_x\delta/2<kT$ (where $F_x$ is the projection of force $\mathbf{F}$ on the line of the thread, and $\delta$ is the distance between the minima of potential relief along the line of the thread), the relative motion of walls along the line of the thread constitutes a drift and is described by the Fokker–Planck equation

$$\frac{\partial n}{\partial t} = D\frac{\partial^2 n}{\partial x^2} + \frac{\partial n}{\partial x} BF_x \qquad (8)$$

where $x$ is the coordinate specifying the relative wall arrangement along the line of the thread, $n(x,t)$. is the probability distribution function for the relative wall arrangement, $D$ is the diffusion coefficient:

$$D = \frac{1}{2}\Omega\delta^2 \exp\left(-\frac{U_1}{kT}\right), \qquad (9)$$

$B$ is the mobility:

$$D = \frac{\Omega\delta^2}{2kT}\exp\left(-\frac{U_1}{kT}\right), \qquad (10)$$

and $\Omega$ is the pre-exponential factor in the Arrhenius formula for the frequency of wall jumps between the equivalent potential relief minima located in the line of the thread (by the order of magnitude, $\Omega$ is equal to the average frequency of the relative vibrations of the walls along the line of the thread). We note that the Einstein relation $D=kTB$ is fulfilled.

The operating mode of NEMSs reliant on the relative motion of nanotube walls, when this motion is described by the Fokker–Planck equation, will be termed the stochastic mode [28–30]. A steady directional wall motion in the Fokker–Planck operating mode is possible when the average distance $x_{dr}=BF_xt$, which the wall travels along the line of the thread in a time $t$ by virtue of the drift under the action of force $\mathbf{F}$, exceeds the distance $x_{di} = \sqrt{2Dt}$ by which it shifts along this line due to diffusion:

$$BF_xt \gg \sqrt{2Dt}. \qquad (11)$$

We substitute the Einstein relation $D=kTB$ into condition (11) to arrive at the conclusion that it is worthwhile employing the stochastic mode in NEMSs for an operating time

$$t > \frac{2k^2T^2}{F_x^2 D} \qquad (12)$$

and wall drift displacements $x_{dr}$ along the line of the thread:

$$x \gg \frac{2kT}{F_x}, \tag{13}$$

On substituting into condition (13) the strongest force $F_x \approx 2kt/\delta$ for which the relative motion of walls is adequately described by the Fokker–Planck equation, we conclude that it is expedient to use the stochastic mode in NEMSs for $x_{dr} \gg \delta$, i.e., for several dozen relative wall jumps between the minima of the potential relief along the line of the thread. Such a wall displacement corresponds to about one turn about the nanotube axis and to a displacement along its axis by several nanometers. The stochastic mode may be employed in some of the NEMSs considered in Section 5, like nanomotors and nanoactuators.

For forces satisfying the condition $F_x \delta / 2 \gg kT$, the stochastic component in relative motion of walls may be neglected. In this case, the motion of the walls is accelerated and is rather accurately described by the equation of motion. The NEMS operating mode which is based on the relative motion of nanotube walls and corresponds to such forces will be referred to as dynamic. This mode allows controllable relative displacement of the walls by any distance along the line of the thread. This operating mode may be employed, for instance, in variable nanoresistors.

The force **F** which sets the movable wall in motion relative to the fixed one may be resolved into two components: the force directed along the nanotube axis (a force of the 1st type), and the force aligned with the tangent to the circumference of the movable wall (a force of the 2nd type). When the potential relief corrugation has no effect on the relative motion of walls, the type of relative motion of walls corresponds to the type of forces applied. Specifically, type-1 forces give rise to the relative wall sliding along the nanotube axis, while type-2 forces lead to the relative rotation of the walls. However, when the potential relief corrugation exerts a significant effect on the relative motion of walls, this correspondence is generally absent. In particular, for a potential relief of the thread type the relative motion of walls along the helical line of the thread may be caused by the action of forces of both types (or their superposition) applied to the movable wall. The relative motion of walls along the helical line of the thread is simultaneously both the relative rotation and the relative sliding along the nanotube axis. Consequently, the type-1 force gives rise not only to the relative sliding of the walls along the nanotube axis, but also to their relative rotation, and the type-2 force effects not only the relative rotation of the walls, but also their relative axial sliding. In the former case, it is possible to create NEMSs operating like a whirligig. This way of setting the walls in relative rotation may be employed in the nanoactuators described in Section 5.6, which consist of a nanobolt–nanonut pair and nanotube-based nanobearings. In the latter case, it is possible to fabricate NEMSs operating like a faucet in which the rotation of a knob is transformed to the translational motion of the bar.

### 4.2 Ways of controlling the nanoelectromechanical systems

Control of NEMSs based on the relative motion of nanotube walls may be effected with the help of (i) a nanomanipulator, (ii) the pressure of a gas being heated and confined between the movable and fixed walls, (iii) a magnetic field, and (iv) an electric field.

The feasibility of controlling the relative motion of nanotube walls by way of a nanomanipulator attached to an atomic-force microscope has been demonstrated experimentally [5, 6]. The drawback of this method of controlling NEMS motion is the large dimensions of the nanomanipulator and the nanomanipulator motion control system in comparison with the dimensions of the NEMS itself.

The feasibility of relative motion of nanotube walls under the pressure of the gas being heated and confined between the movable and fixed walls was shown by simulations using the molecular dynamics method [78]. In Ref. [78] it was suggested that the gas inside a NEMS should be rapidly heated by a pulsed laser. However, this means of control cannot be used for continuous NEMS operation, because the NEMS has to be cooled after each single actuation of the movable wall.

When the movable wall is metallic or metal-filled and the remaining walls are semiconducting, it was suggested that the movable wall be driven by a nonuniform magnetic field [79]. When such a wall moves at a velocity $V$ in a nonuniform magnetic field $B$, the variation of the magnetic flux $\Phi$ induces in the wall a current

$$i = -G\frac{d\Phi}{dt} = -GS\frac{dB}{dt} = -GSV\frac{dB}{dz}, \tag{14}$$

where $G$ is the conductivity, and $S$ is the wall cross-section area. The wall energy $U_m$ in the magnetic field is defined by the expression

$$U_m = -\mu B = -SiB = -GS^2VB\frac{dB}{dz}, \quad (15)$$

where $\mu=Si$ is the magnetic dipole moment of the wall. The force $F_m$ exerted by the magnetic field on the moving wall is expressed in the form

$$F_m = \frac{dU_m}{dz} = GS^2V\left[\left(\frac{dB}{dz}\right)^2 + B\frac{d^2B}{dz^2}\right]. \quad (16)$$

However, the experiments, simulations, and numerical estimates proving the feasibility of this way of controlling nanotube wall motion are utterly lacking to date.

The motion of the inner wall filled with a magnetic material may be controlled with the aid of a magnetic field. Several methods have been elaborated for the fabrication of MWNTs containing magnetic materials inside of them [80–82]. However, the walls of nanotubes with magnetic materials, fabricated by these methods, contain too many defects, which makes unlikely the relative motion of such walls.

To control the motion of the movable wall by means of an electric field, it is required to charge this wall or produce a wall with an electric dipole moment. It has been hypothesized that the movable wall may be charged as a result of doping [18, 19]. According to calculations, the valence electron of an isolated potassium atom in the metallofullerene K@$C_{60}$ is completely transferred to fullerene $C_{60}$ [83]. A similar charge transfer from metal atoms to a nanotube also takes place when metallofullerenes reside inside a single-walled nanotube [84]. Further charge transfer to the electrodes connected to the nanotube may occur in a nanoelectromechanical system based on such nanotubes. In this way, the metal atoms confined in the nanotube may become ions, and the internal wall containing the metal atoms may be set in motion with the help of an electric field [85]. The technology for the fabrication of these walls was proposed in Ref. [85]; it is based on the annealing of a nanotube with metallofullerenes inside of it and makes it possible to obtain only very short walls of uncontrollable length [86].

According to calculations, the electrostatic potentials at the open and capped ends of a SWNT are significantly different [87]. Consequently, if a wall is made open at one end and capped at the other, it would possess an electric dipole moment. Charge distribution simulations have revealed the emergence of a dipole moment in the chemical adsorption of $Br_2$ [88] and $H_2O$ [89] molecules at the edge of a single-walled nanotube. Furthermore, the sign of the charge transferred to the edge depends on what atoms (donors or acceptors of electrons) are adsorbed at the open nanotube edges. We therefore suggest that the electric dipole moment of the wall be increased by way of adsorption of the donors and acceptors of charge at the opposite open edges of the wall. The motion of this wall may be controlled with the aid of a nonuniform electric field.

Electric dipole moments were calculated by the semiempirical molecular orbital method by the example of a 3.15-nm long SWNT (5,5) in two cases of chemical modification of the nanotube ends [90]. In the first case, one nanotube end was capped and electron donors were adsorbed at all dangling bonds of the open end. In the second case, both nanotube ends were open, with electron donors adsorbed at all dangling bonds of one end, and electron acceptors at all dangling bonds of the other end. Hydrogen and fluorine atoms were selected as the electron donors and acceptors, respectively. Our calculations showed that the electric dipole moments of the walls are equal to $4.536\times10^{-29}$ C m in the former case, and $7.397\times10^{-29}$ C m in the latter. The feasibility of controlling by means of a nonuniform electric field the motion of a NEMS based on a nanotube comprising walls-dipoles was demonstrated by the example of a gigahertz oscillator [90] (see Section 5.5).

# 5. Nanoelectromechanical systems reliant on the motion of nanotube walls

## 5.1 Nanobearings and nanomotors

The idea that MWNTs may be ideal nanobearings completely free from wear was put forward shortly after the discovery of nanotubes [91]. The energy dissipation during operation of DWNT-based nanobearings was simulated by the molecular dynamics method. By way of example, for nanobearings based on the DWNTs (9,9)@(14,14) and (9,9)@(22,4) with an interwall spacing of about 3.4 Å at a temperature of 300 K and an angular velocity of $3.1\times10^{11}$ s$^{-1}$, the energy dissipation equals 3.3 and 3.7 meV per atom, respectively [92]. The energy dissipation rate was found to increase with temperature and to decrease with the spacing between the DWNT walls [13]. The spectra of the nanobearings exhibited low-frequency longitudinal, breathing, and torsional wall oscillations, medium-frequency oscillations which correspond to variations of the angles between the bonds, and high-frequency oscillations which correspond to variations of the bond lengths [13]. All these oscillations make contributions to the energy dissipation.

MWNT-based nanobearings have been used in recently fabricated nanomotors [26, 27]. In both nanomotors, a metal plate (the rotor) 200–500 nm in size was attached to the center of a movable MWNT wall secured between electrodes (Fig. 3). The control electrodes of the stator were arranged around the MWNT. The capacitance of this system, and hence the corresponding electrostatic energy, depends on the rotation angle of the plate attached to the outer MWNT wall. That is why the voltage applied across the plate and the stator electrodes produces an electrostatic torque which acts on the movable wall, thereby making it rotate.

In the former nanomotor realized, the plate was attached to the outer MWNT wall [26]. To open up the possibility of rotating this wall, a part of it in the region between the rotor electrodes and the plate was broken by applying an electrostatic force to the plate. In this nanomotor design, the inner walls make up the stator, and the outer wall serves as the rotor. To set the rotor in motion, the voltages $V_1=V_0\sin(\omega t)$, $V_2=V_0 \sin(\omega t-\pi)$, $V_3=V_0\sin(2\omega t+p/2)$, and $V_r=-V_0$ were applied respectively to the three electrodes of the stator and to the rotor.

More recently, a nanomotor with improved design was made [26, 27]. In the fabrication of this motor, 5–10 outer MWNT walls in the middle of the region between the rotor electrodes were removed by ohmic heating. After that, the metal plane was attached to the middle of the outer remaining wall. In this nanomotor design, the inner walls make up the rotor, and the portions of the outer walls that survive at the nanotube ends serve as the stator. In this case, the caps of the outer walls (removed in the middle of the nanotube) hinder the displacement of the movable wall along the nanomotors axis.

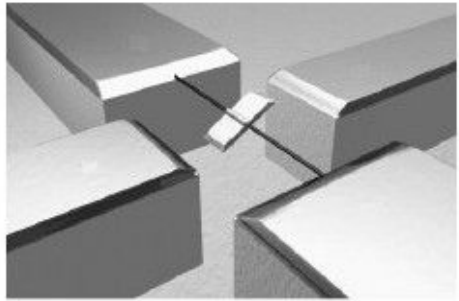

Figure 3. Schematics of a nanomotor reliant on the relative rotation of the walls of a carbon nanotube [26].

**5.2 Nanorelay and memory cell**

A suggestion for a nanorelay based on the relative motion of carbon nanotube walls was recently put forward [16, 17]. Figure 4 shows several possible nanorelay schemes. In these nanorelays, a DWNT is attached to electrode *3*. The movable inner DWNT wall *1* telescopically extends from the outer wall *2* under the action of an electrostatic force $F_e$ and is attracted to electrode *4* by the van der Waals force $F_a$ due to the interaction between the inner wall and electrode *4* (turning the relay on). The inner wall is pulled back into the outer one by a van der Waals force $F_W$ due to the interaction between the inner and

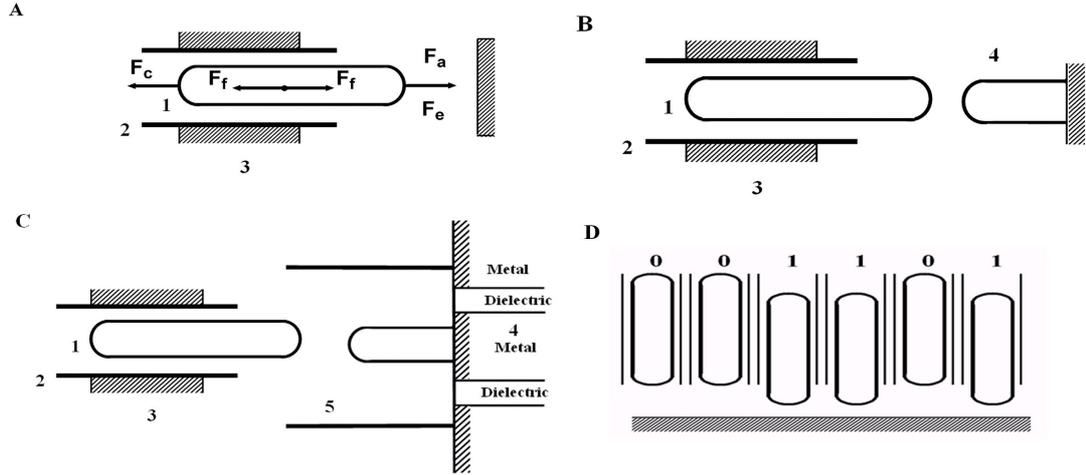

Figure 4. Schematics of DWNT-based electromechanical nanorelays (in the 'on' position): (a) a nanorelay with the second planar metallic electrode, (b) a nanorelay with a carbon nanotube as the second electrode, and (c) a nanorelay with a carbon nanotube as the control electrode: 1 — movable inner wall, 2 — fixed outer wall, 3 and 4 — electrodes. (d) Example of a storage made of DWNT-based memory cells.

outer walls (turning the relay off). The second electrode may be either metallic (Fig. 4a) [16] or a single-walled carbon nanotube (Fig. 4b) [17]. A nanorelay design with the third, control electrode made of a single-walled carbon nanotube has also been put forward (Fig. 4c) [17]

As indicated in Section 3, when the inner movable wall of the first electrode is nonchiral and commensurate with the outer wall, in the balance of forces that determine the nanorelay operation it is necessary also to take into account the static friction force $F_f$ which impedes the relative motion of walls. The force $F_f$ may be neglected for the relative motion of incommensurate and commensurate chiral wall pairs of the first electrode. The condition which permits the system to operate as a relay is bistability — the existence of two potential energy minima. An analysis of the balance of forces during the nanorelay operation shows that $F_a > F_W$ for the nanorelay with the second metallic electrode (Fig. 4a), and hence both potential energy minima exist in the absence of the control voltage [16, 17]. This nanorelay may be employed as an *nonvolatile memory cell* for any pair of walls of the first electrode. A memory device consisting of such cells is diagrammed in Fig. 4d. For a nanorelay based on the incommensurate DWNT (8,2)@(12,8) with the scheme shown in Fig. 4a, the turn-on and turn-off voltages (±3 V) and the switching time ($10^{-11}$ s) were calculated, which corresponds to an operating frequency of 100 GHz [16].

A nanorelay in which both electrodes are made of carbon nanotubes offers several advantages over a nanorelay in which the second electrode is a metallic nanowire. The use of nanotubes as electrodes permits, first, decreasing the nanorelay dimensions and, second, attaining a higher evenness in dimensions and characteristics of the electrode surfaces. For a nanorelay with the second nanotube electrode, one obtains $F_a < F_W$. Consequently, for incommensurate and commensurate chiral wall pairs of the first electrode (in the absence of the friction force $F_f$), the nanorelay may stay in the 'on' position only when the control voltage is applied [17]. This nanorelay offers promise as a random access memory (RAM) cell. According to estimates, for a nanorelay with the second and control electrodes made of nanotubes and having the smallest possible radii, the turn-on voltage and the voltage required for the retention in the 'on' state are equal to 6 and 4.8 V, respectively [17].

For a DWNT with commensurate nonchiral walls in which the friction force $F_f$ is significant, it is possible to select the length of the outer wall so that the condition $F_a + F_f > F_W$ is fulfilled. In this case, the nanorelay with the nanotube second electrode may also be used as a read-only memory (ROM) cell. According to estimates, a nanorelay with the nanotube second electrode and the DWNT (5,5)@(10,10)-based first electrode may be bistable in the absence of control voltage when the wall overlap length lov of the DWNT (5,5)@(10,10) exceeds the critical value $l_c$=4 nm [17]. However, the bistability condition $l_{ov} > l_c$ would not suffice for the adequate nanorelay operation as an energy-independent memory cell: the potential barrier $\Delta U$ which separates the 'on' and 'off' positions may be overcome due to the thermal diffusion of the inner wall. The probability $p$ of such an event is defined by the Arrhenius formula $p = \Omega \exp(-\Delta U / kT)$, where $\Omega$ is the pre-exponential factor. According to simulations of the reorientation dynamics of the $C_{60}@C_{240}$ nanoparticle shells by the molecular dynamics method, the pre-exponential factor $\Omega$ exceeds the frequency of small oscillations about the equilibrium position by one order of magnitude [93, 94]. Figure 5 depicts the dependence of the 'on'-position lifetime $t=p^{-1}$ on the wall overlap length $l_{ov}$, calculated for a nanorelay with the nanotube second electrode and the DWNT (5,5)@(10,10)-based first electrode [17]. This plot demonstrates the feasibility of utilizing the nanorelay with the nanotube second electrode as a nonvola-tile memory cell.

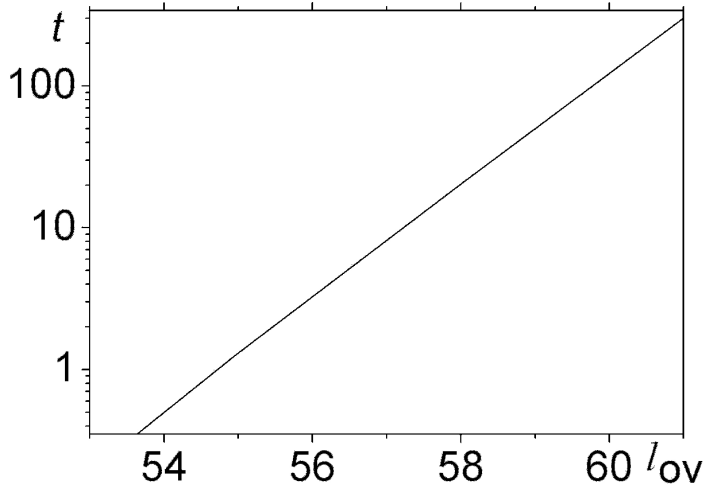

**Figure 5**. Dependence of the 'on'-position lifetime $t$ on the wall overlap length $l_{ov}$ at a temperature of 300 K, calculated for a nanorelay with the (5,5) nanotube second electrode and the DWNT (5,5)@(10,10)-based first electrode.

### 5.3 Nanoresistors

According to calculations, the conductivity of several carbon nanosystems, like a fullerene placed between two SWNTs [95] and a DWNT with telescopically extended walls [34–37], may alter by several orders of magnitude under relative displacements of the parts of a nanosystem by a distance of about 1 Å We suggest that the relative displacements of the parts of both of these nanosystems and hence the nanoresistor conductivity be controlled with the aid of a DWNT-based nanoactuator [28–30]. In particular, advantage can be taken of a nanobolt–nanonut pair to transform an axially directed force into the rotation of the conducting wall or a torque-producing force into the axial sliding of this wall. The schematics of nanoresistors based on the above nanosystems are given in Fig. 6.

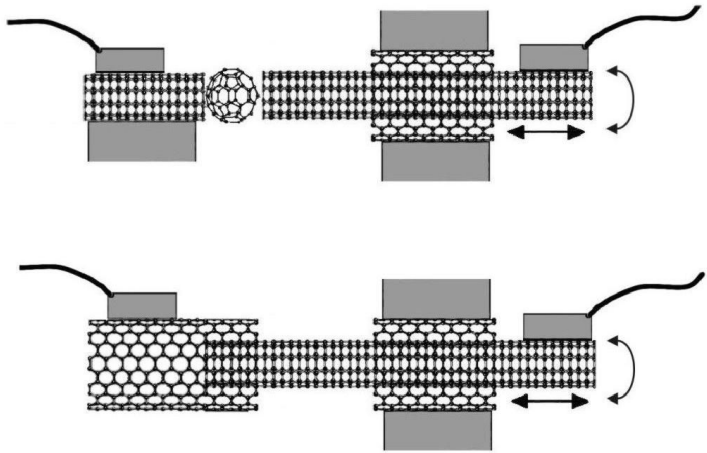

Figure 6. Basic diagrams of nanoresistors based on carbon nanotubes.

### 5.4 Electromechanical nanothermometer

We have come up with the conceptual idea of an electro-mechanical nanothermometer reliant on the interaction and relative motion of carbon nanotube walls, where the temperature is taken by measuring the conductivity [17, 39]. The relative position of the nanotube walls varies under temperature fluctuations as a result of relative thermal wall vibrations. The probability $p(z)$ of the relative displacement of the walls by a distance $z$ along the nanotube axis depends on the interwall interaction energy $U(z)$ as

$p(z) \sim \exp(-U(z)/kT)$. The total nanotube conductivity at a temperature $T$ with the inclusion of thermal wall vibrations is defined by the following expression:

$$G_{tot}(T) = \frac{\int_{-\infty}^{\infty} G(z,T) \exp(-U(z)/kT) dz}{\int_{-\infty}^{\infty} \exp(-U(z)/kT) dz}. \quad (17)$$

The following conditions should be fulfilled for the operation of an electromechanical nanothermometer reliant on the interaction of carbon nanotube walls: $G(z,T)$ depends only slightly on the temperature for a fixed relative position of the walls (condition A); the conductivity $G(z,T)$ essentially depends on the z-coordinate which specifies the relative position of the walls (condition B); the characteristic amplitude of thermal wall vibrations is high enough for the thermal vibrations to make the main contribution to the temperature dependence of the conductivity (condition C) and is low enough for the relative wall displacements not to disrupt the nanothermometer operation (condition D).

Possible schemes of an electromechanical nanothermometer are depicted in Fig. 7. We considered the realizability of this nanothermometer by way of the example of a DWNT (6,6)@(11,11). The DWNT (6,6)@(11,11) conductivity depends essentially on the relative position of the walls for both nanothermometer schemes given in Fig. 7 [35]. Condition B is therefore fulfilled. Experimental data suggest that the conductivity of single-walled nanotubes possesses only a weak dependence on the temperature for $T >$ 50 K [97]. Calculations show that the DWNT conductivity depends only slightly on the temperature for $T$ > 100 K [39, 96]. Therefore, condition A is also fulfilled.

The dependences of the conductivity $G(z)$ and the interwall interaction energy $U(z)$ on the relative displacement $z$ of the walls are defined by the symmetry of a DWNT (6,6)@(11,11). These dependences have common properties: their periods and extrema coincide. Furthermore, due to the incompatibility of wall rotation symmetries, both the conductivity and the interaction energy of the walls depend only slightly on the angle of relative wall rotation. Since adequate nanothermometer operation implies the absence of diffusive motion of the short wall relative to the long one, integration in formula (17) may be performed only in the vicinity of the minimum of the interwall interaction energy. In this domain, the dependence (7) of the interwall interaction energy on the relative wall position can be approximated by the expression

$$U(z') = U_1 + \frac{\pi \Delta U_z}{b^2} z'^2, \quad (18)$$

where $z'$ is the displacement of the movable wall with respect to the position corresponding to the minimum of the interwall interaction energy, and $U_1$ is the value of the wall interaction energy corresponding to the minimum. Since the extreme points of the dependences of the interwall interaction energy $U(z')$ and the conductivity $G(z')$ coincide, we approximated the dependence $G(z')$ for small displacements $z'$ as follows:

$$G(z') = G_1(T)(1 + \gamma z'^2), \quad (19)$$

where $G_1$ is the conductivity value corresponding to the minimum of the interwall interaction energy. The $\gamma$-parameter value estimated from the dependence $G(z')$ given in Fig. 3 in Ref. [35] is $\gamma \sim 850 \text{Å}^{-2}$ for a wall overlap of 2.45 nm and the nanothermometer scheme depicted in Fig. 7a.

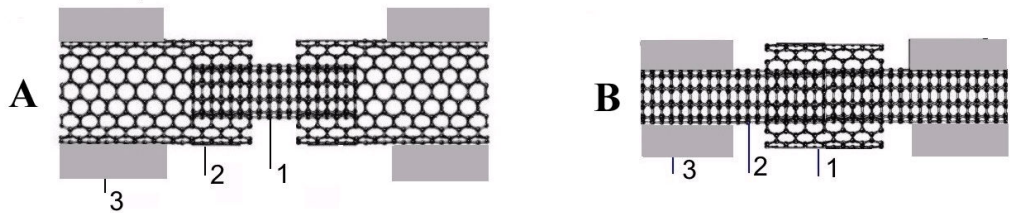

Figure 7. Schematics of electromechanical DWNT-based nanothermometers: (a) a telescopic nanothermometer with a movable inner wall, and (b) a nanothermometer with a movable shuttle from the outer wall: 1 — movable wall, 2 — fixed wall, and 3 — electrodes.

On substituting expressions (18) and (19) into formula (17) we obtained the following relation for the temperature dependence of nanothermometer conductivity:

$$G(T) = G_1(T)\left(1 + \frac{\gamma b^2 kT}{\pi \Delta U_z}\right) = G_1(T)(1 + HT), \quad (20)$$

According to the data calculated by the method of density functional, the barrier height is $\Delta U_z$=1.96 eV for a wall overlap of 2.45 nm [12]. Therefore, $HT$≈140 for a temperature $T$=300 K, which signifies the fulfillment of the condition C.

The condition D is fulfilled when the movable wall displacement $d$ arising from diffusion turns out to be shorter than the movable wall–electrode spacing $L_e$:

$$d = \sqrt{2Dt} < L_e, \qquad (21)$$

where $D$ is the diffusion coefficient for the movable wall, and $t$ is the nanothermometer operation time. The diffusion coefficients for the relative motion of DWNT walls were recently calculated with the help of the method of density functional [98]. By taking advantage of these data for the DWNT (6,6)@(11,11), we arrived at the conclusion that the interelectrode distance for a nanothermometer that would operate for 100 years without failure is equal to about 40 nm for both schemes depicted in Fig. 7. Therefore, the length of the proposed nanothermometer is actually determined by the length required to connect the electrodes to the nanotube, and it may be employed for nanolocal temperature measurements. The nanothermometer may be calibrated with the aid of a thermocouple, and it is possible to achieve the same measurement accuracy as for the thermocouple, since the nanothermometer relies on conductivity measurements (the accuracy of room-temperature measurements for copper–constantan and chromel–alumel thermocouples amounts to 0.1 K).

### 5.5 Gigahertz oscillator

On the basis of an MWNT it is possible to make a nanomechanical device in which there occurs a periodic motion at a very high frequency – on the order of 1–100 GHz [78, 85]. It has been suggested that these devices, which are termed gigahertz oscillators, be employed, in particular, in nanoactuators [18, 19]. The design and operation of the gigahertz oscillator are schematized in Fig. 8.

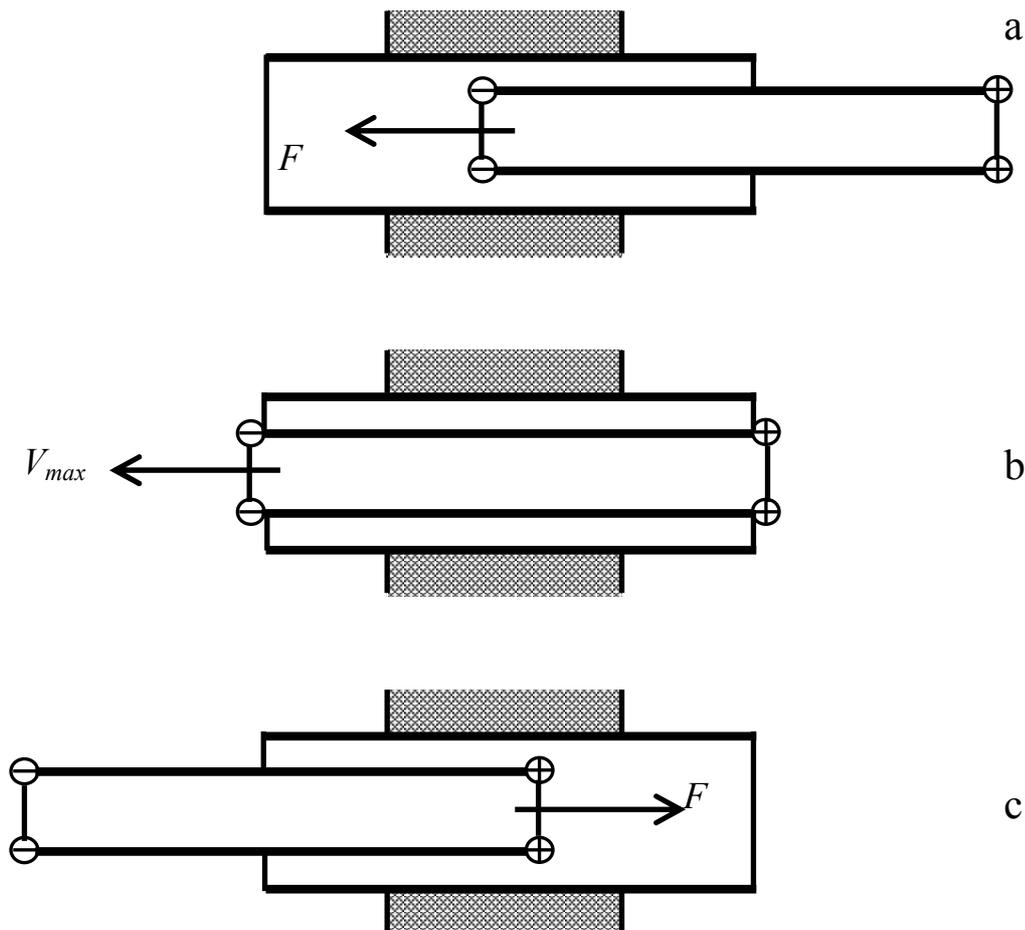

Figure 8. Sequential positions of the walls of a gigahertz oscillator during half the oscillation period: (a) and (c) the maximum telescopic extension of the inner wall; the van der Waals force $F_W$ retracts the inner wall into the outer one; (b) moving under its own momentum, the inner wall passes at the highest velocity $V_{max}$ through the position with the minimal potential energy.

The van der Waals force FW, which retracts the telescopically extended inner wall into the outer one, essentially depends on the relative position of the walls only when the separation of the wall edges is on the order of the interwall distance. Therefore, in the standard theory of the gigahertz oscillator, the dependence of the force FW on the distance x between the wall centers may be approximated by the following expression [18, 19]:

$$F_W(x) = \begin{cases} F_W, & |x| > |L-l|/2, \\ 0, & |x| < |L-l|/2, \\ 0, & |x| > |L+l|/2, \end{cases} \quad (22)$$

where $L$ and $l$ are the outer and inner wall lengths, respectively. For such an oscillator model, the dependence of the potential energy on the relative wall position is of the form given in Fig. 9. The natural frequency of small oscillations is absent from systems with such a potential, while the frequency of high-amplitude oscillations (which are targeted for use in NEMSs) depends strongly on the amplitude. That is why it is required to maintain a constant amplitude of these oscillations for the nanotube-based gigahertz oscillator frequency to remain invariable. The oscillation period $T$ is defined by the expression

$$T = 2(t_{in} + t_{out}) = 2\left[(L-l) + 4s\right]\sqrt{\frac{m}{2F_W s}}, \quad (23)$$

where m is the inner wall mass, $s$ is the maximum distance by which the inner wall extends from the outer one, $t_{in} = (L-l)\sqrt{m/2F_W s}$ is the time it takes for the inner wall to travel from one end of the outer wall to the other, and $t_{out} = 2\sqrt{2ms/F_W}$ is the time of an inner wall motion during which one of its ends is beyond the outer wall.

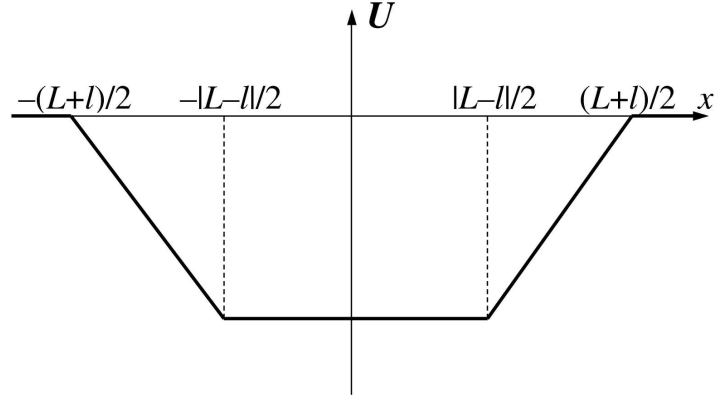

Figure 9. Schematic representation of the oscillator potential energy $U$ as a function of the distance $x$ between wall centers.

The characteristics of gigahertz oscillators were comprehensively studied by simulations using the molecular dynamics method [79, 99–106]. It was shown, in particular, that dynamic friction is responsible for oscillation energy dissipation: their free oscillations executed with a frequency which increases with time and decay in a time on the order of several nanoseconds [99, 100, 105] with a quality factor $Q=\Delta E/E \sim 100$–1000, where $E$ and $\Delta E$ are the oscillator energy and the energy loss during one period [104, 105]. An analysis of the results of damped oscillation simulations revealed that the dependence of the dynamic friction force on the inner wall velocity $V$ may be approximated as $F= -aV-b\mathrm{sign}(V)V^2$ [106]. Then, during one period the friction force does the work

$$A_f = -2aV_{max}^2\left(\frac{t_{out}}{3} + t_{in}\right) - 2bV_{max}^3\left(\frac{t_{out}}{4} + t_{in}\right), \quad (24)$$

where $V_{max} = \sqrt{2F_W s/m}$ is the highest velocity of the inner wall [90].

The possibility of compensating for the energy dissipation was considered in the case when the motion of the inner oscillator wall is controlled by an external harmonic force $F(t)=F_0 \cos\omega t$ aligned with the wall axis, with $\omega$ corresponding to the desired oscillation frequency [90]. The work $A_c$ of the force $F(t)$ in a period $T$ is defined by the expression

$$A_c = \frac{4F_W F_0}{m\omega^2} \cos\left(\frac{\omega t_{in}}{2}\right), \quad (25)$$

where $\omega=2\pi/T$ is the circular oscillation frequency. During the same time, the friction force does the work $A_f = -F_W s/Q$. From the condition $A_c + A_f = 0$, which signifies that the work of the controlling force compensates for the energy dissipation, we find the controlling force amplitude $F_0$:

$$F_0 = \frac{ms\omega^2}{4Q\cos(\omega t_{in}/2)}. \qquad (26)$$

An analysis of expression (26) reveals that $F_0$ is minimal for the walls of equal length, $L=l$. In the case of equal wall lengths, the oscillation frequency is expressed as $\omega = (\pi/2)\sqrt{F_W/2ms}$ and relation (24) for the work done by the friction force takes on the form

$$A_f = -\frac{8a\sqrt{2F_W s}s}{3\sqrt{m}} - \frac{4bF_W s^2}{m}, \qquad (27)$$

While expression (26) for the controlling force amplitude reduces to

$$F_0 = \frac{\pi^2 F_W}{32Q}. \qquad (28)$$

The operating characteristics of a gigahertz oscillator were estimated by the example of an oscillator based on the DWNT (5,5)@(10,10) with equal wall lengths $L=l=3.15$ nm (the frequency of oscillations with an amplitude $s=1$ nm for this oscillator amounts to about 400 GHz). According to *ab initio* calculations, the interwall interaction energy in the DWNT (5,5)@(10,10) is 23.83 meV per atom of the outer wall [12], which corresponds to the value of $F_W = 625$ pN for the force drawing in the inner wall. For this value of the force $F_W$ and a quality factor $Q=1000$, the amplitude of controlling force reaches $F_0 = 0.193$ pN.

In particular, the harmonic controlling force may be applied to a wall-dipole with the aid of a nonuniform electric field. The calculated results for the dipole moments of wallsdipoles, given in Section 4.2, were used for estimating the voltage required to produce such a nonuniform electric field which would control the wall-dipole motion and maintain the stable oscillator frequency [90]. It was found that for a spherical capacitor with plates approximately 100 nm in diameter and a plate separation of about 10 nm it is possible to control the wall-dipole motion by applying the voltage $U(t)=U_0 \cos\omega t$ with the amplitudes $U_0 = 2.24$ and 2.77 V in the first and second cases of the walls-dipoles described in Section 4.2.

The master voltage was estimated for a nonuniform electric field produced by a charge distributed over the surface with a relatively weak curvature. Using the field produced by a charge on the surface with a stronger curvature, in particular, the field near a tip, would enable the control of a wall-dipole motion by applying a significantly lower voltage.

### 5.6 Nanoactuator

As indicated in Section 3, when the dependence of the interaction energy $U(\phi,z)$ of walls on their relative position possesses helical symmetry, the walls may be utilized as a nanobolt–nanonut pair. We suggest the use of this pair in a nanoactuator (Fig. 10) intended to transform into wall rotation the translational force aligned with the nanotube axis [17, 32, 33]. In such a nanoactuator, the stationary fixed nanotube wall 1 makes up the stator. Walls *1* and *2* are a rotation nanobearing, i.e., the condition $E_\phi \ll E_z$ is fulfilled for them, where $E_\phi$ and $E_z$ are the potential barriers for the relative rotation of these walls and their relative sliding along the nanotube axis, respectively. This condition is fulfilled when walls 1 and 2 are nonchiral and commensurate [making a DWNT of the type $(n,n)@(m,m)$ or $(n,0)@(m,0)$]. For adequate nanoactuator operation it is desirable that the relative position of walls *2* and *3* be fixed. The relative displacement of walls *2* and *3* along the nanotube axis will be prevented when wall *3* is nonchiral and commensurate with wall *2*. To avoid the relative rotation of walls *2* and *3* we suggest that atomic structural defects be made in wall *3*. The same defects periodically located at identical sites of the unit cells of wall *3* may be used to produce a nanobolt–nanonut pair from walls *3* and *4*. This pair serves to transform into rotor rotation the force applied to wall *4* and aligned with the nanotube axis.

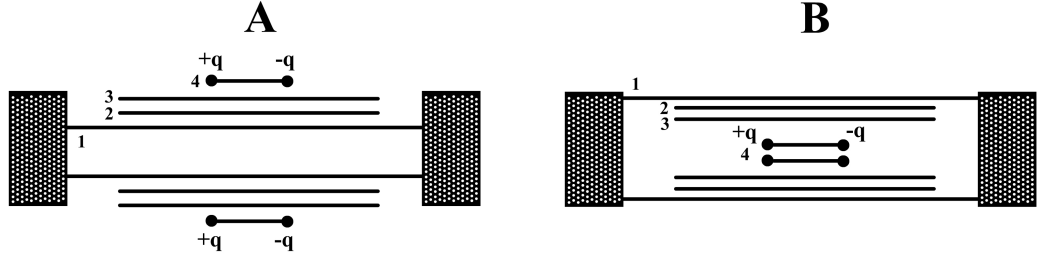

Figure 10 Schematics of a nanoactuator: (a) with the inner wall as the stator; (b) with the outer wall as the stator. Fixed wall *1* — the stator. Jointly rotating walls *2* and *3* — the rotor. Walls *3* and *4* make up a nanobolt — nanonut pair. The charges at the edges of wall *4* may be obtained by chemical adsorption and employed to control the nanoactuator with the aid of an electric field.

Wall *4* will come into helical motion relative to wall *3* provided the following conditions are fulfilled: (i) the initial kinetic energy corresponding to the motion along the line of the thread is higher than the potential barrier $E_1$ for this motion:

$$\frac{M\mathbf{V}^2 \sin^2 \chi}{2} > E_1, \quad (29)$$

where $M$ and $\mathbf{F}$ are the mass and velocity of wall *4*, and $\chi$ is the thread angle; (ii) the initial kinetic energy corresponding to the motion across the line of the thread is lower than the potential barrier $E_2$ for twisting-off the thread:

$$\frac{M\mathbf{V}^2 \cos^2 \chi}{2} < E_2. \quad (30)$$

It is easily shown that conditions (29) and (30) are simultaneously fulfilled only when

$$\cot^2 \chi < \frac{E_2}{E_1} = \beta. \quad (31)$$

Condition (31) assesses the feasibility of making a nanoactuator on the basis of a given nanobolt–nanonut pair. Note that it follows from condition (31) that nanobolt–nanonut pairs with a small relative depth of thread are suited for making a nanoactuator, when the thread angle exceeds 45°.

The conservation of the angular momentum of a nanoactuator determines the minimal rotor length $L_3$, whereby in the given nanoactuator it is possible to turn the rotor by an angle $\xi$:

$$L_3^{\min} = \frac{L_4^2 R_4^2}{L_4 R_4^2 + \xi(R_2^3 + R_3^3)\sin \chi}, \quad (32)$$

where $L_4$ is the length of wall *4*, and $R_2$, $R_3$, and $R_4$ are the respective radii of walls *2*, *3*, and *4*.

The complete list of four-walled nanotubes which may be employed in the making of this nanoactuator was given in Ref. [33]; also considered in Ref. [33] was a nanoactuator based on the (11,2)@(12,12)@(19,19)@(26,26) nanotube. The (11,2)@(12,12) nanobolt–nanonut pair, which makes up the inner walls in this nanoactuator, has a rather large relative depth of thread: $\beta$=5.812. Furthermore, condition (31) is fulfilled for any relative depth of thread for a thread angle $\chi$=70.9° calculated for this nanobolt–nanonut pair.

## 6. Summary

In conclusion, let us list the results achieved in the development of nanotechnology techniques required for the fabrication of nanotube-based nanoelectromechanical systems. At the present time, with the help of a nanomanipulator it is possible to transfer SWNTs [107, 108] and set the walls of multi-walled nanotubes in relative motion [5]. Also effected was the removal of caps which covered the ends of nanotubes [109–111] and cutting nanotube walls into parts of desired lengths [112]. Several nanoelectromechanical systems considered in our report, in particular, a nanothermometer and a nanoactuator, can be made only of walls with specific chirality indices. In this connection, we note that a technique has been devised recently for the unambiguous determination of wall chirality indices [113].

The majority of the nanoelectromechanical systems that were the concern of our report were considered by way of the example of double-walled nanotubes. These nanotubes may be obtained by different synthesis techniques: in a conventional arc for the production of carbon nanotubes [7, 114], in the same arc in the presence of hydrogen and a catalyst [115, 116], and in the catalytic decomposition of hy-drocarbons [117–123], as well as by heating [86, 124, 125] or electron irradiation [126] of single-walled nanotubes containing a fullerene chain inside.

Therefore, great progress in nanotechnology gives us hope that the carbon nanotube-based nanoelectromechanical systems considered in our report may be fabricated in the near future.